\documentstyle[11pt]{article}
\def\LaTeX{L\kern-.25em\raise.425ex\hbox{a}\kern-.075em\TeX}

\def\ni{\noindent }
\def\eq #1{Eq.(\ref{#1})}
\def\l{\left}
\def\r{\right}
\def\pa{\partial}
\def\fr{\frac}
\def\se #1{sec. \ref{#1}}

\def\nth{${\rm n}^{{\rm th\;}}$}

\def\y1{\mbox{$y'$}}
\def\yt{\mbox{$y''$}}

\def\yk{\mbox{$y^{(k)}$}}
\def\yn{\mbox{$y^{(n)}$}}
\def\ym{\mbox{$y^{(n-1)}$}}
\def\ymm{\mbox{$y^{(n-2)}$}}
\def\F{{\cal F}}
\def\G{{\cal G}}
\def\H{{\cal H}}

\def\mut{\stackrel{\sim}{\mu}\!}

\def\bigl({\left({\vrule height1.29em width0em depth1.29em}}
\def\bigr){{\vrule height1.29em width0em depth1.29em}\right)}
\def\medl({\left({\vrule height1em width0em depth1em}}
\def\medr){{\vrule height1em width0em depth1em}\right)}
\def\|{\'\i}

\begin{document}

\title{Integrating factors for second order ODEs}
\author{E.S. Cheb-Terrab$^{a,b}$ and A.D. Roche$^a$ }

\date{}
\maketitle
\thispagestyle{empty}

\centerline {\it $^a$Symbolic Computation Group}
\centerline {\it Computer Science Department, Faculty of Mathematics}
\centerline {\it University of Waterloo, Ontario, Canada.}

\medskip
\centerline {\it $^b$Department of Theoretical Physics}
\centerline {\it State University of Rio de Janeiro, Brazil.}

\bigskip
\centerline {\small{ \it {Journal of Symbolic Computation, V. 27, No. 5, pp. 501-519 (1999)}}}
\centerline{ \small {\it{(Original version received 25 November 1997)}}}

\begin{abstract}

A systematic algorithm for building integrating factors of the form
$\mu(x,y)$, $\mu(x,\y1)$ or $\mu(y,\y1)$ for second order ODEs is
presented. The algorithm can determine the existence and explicit form of
the integrating factors themselves without solving any differential
equations, except for a linear ODE in one subcase of the $\mu(x,y)$
problem. Examples of ODEs not having point symmetries are shown to be
solvable using this algorithm. The scheme was implemented in Maple, in the
framework of the {\it ODEtools} package and its ODE-solver. A comparison
between this implementation and other computer algebra ODE-solvers in
tackling non-linear examples from Kamke's book is shown.

\end{abstract}

\section{Introduction}

Although in principle it is always possible to determine whether a given
ODE is exact (a total derivative), there is no known method which is
always successful in making arbitrary ODEs exact. For \nth order ODEs - as
in the case of symmetries - integrating factors ($\mu$) are determined as
solutions of an \nth order linear PDE in $n$+1 variables, and to solve
this determining PDE is a major problem in itself.

Despite the fact that the determining PDE for $\mu$ naturally splits into
a PDE system, the problem is - as a whole - too general, and to solve it a
restriction of the problem in the form of a more concrete ansatz for $\mu$
is required. For example, in a recent work by \cite{bluman1}
the authors explore possible ansatzes depending on the given ODE, which are
useful when this ODE has known symmetries of certain type. In another
work, \cite{wolf} explores the use of computer algebra to try various
ansatzes for $\mu$, no matter the ODE input, but successively increasing
the order of the derivatives (up to the \nth$-1$ order) on which $\mu$
depends; the idea is to try to maximize the splitting so as to increase
the chances of solving the resulting PDE system by first simplifying it
using differential Groebner basis techniques.

Bearing this in mind, this paper presents a method, for second order
explicit ODEs\footnote{We say that a second order ODE is in explicit form
when it appears as $\yt-\Phi(x,y,\y1)=0$. Also, we exclude from the
discussion the case of a linear ODE and an integrating factor of the form
$\mu(x)$, already known to be the solution to the adjoint ODE.}, which
systematically determines the existence and the explicit form of
integrating factors when they depend on only two variables, that is: when
they are of the form $\mu(x,y)$, $\mu(x,\y1)$ or $\mu(y,\y1)$. The
approach works without solving any auxiliary differential equations -
except for a linear ODE in one subcase of the $\mu(x,y)$ problem - and is
based on the use of the forms of the ODE families admitting such
integrating factors. It turns out that with this restriction - $\mu$
depends on only two variables - the use of differential Groebner basis
techniques is not necessary; these integrating factors, when they exist,
can be given directly by identifying the input ODE as a member of one of
various related ODE families.

The exposition is organized as follows. In \se{intfactor}, the standard
formulation of the determination of integrating factors is briefly
reviewed and the method we used for obtaining the aforementioned
integrating factors $\mu(x,y)$, $\mu(x,\y1)$ or $\mu(y,\y1)$ is presented.
In \se{mu_and_X}, some aspects of the integrating factor and symmetry
approaches are discussed, and their complementariness is illustrated with
two ODE families not having point symmetries. Sec. \ref{tests} contains
some statistics concerning the new solving method and the second order
non-linear ODEs found in Kamke's book, as well as a comparison of
performances of some popular computer algebra packages in solving a
related subset of these ODEs. Finally, the conclusions contain some
general remarks about the work.

Aside from this, in the Appendix, a table containing extra information
concerning integrating factors for some of Kamke's ODEs is presented.

\section{Integrating Factors and ODE patterns}
\label{intfactor}

In this paper we use the term ``integrating factor" in connection with the
explicit form of an \nth order ODE

\begin{equation}
% {\frac {d^ny}{dx^n}}=\Phi(x,y,\y1,...,\ym)
\yn-\Phi(x,y,\y1,...,\ym) = 0
\label{oh}
\end{equation}

\ni so that $\mu(x,y,\y1,...,\ym)$ is an integrating factor if

% after multiplying the ODE by $\mu$ we obtain a total derivative

\begin{equation}
% \mu\l({\frac {d^ny}{dx^n}}-\Phi\r) = \frac{d}{dx}R(x,y,\y1,...,\ym)
\mu\l(\yn-\Phi\r) = \frac{d}{dx}R(x,y,\y1,...,\ym)
\label{exact_oh}
\end{equation}

\ni for some function $R$. The knowledge of $\mu$ is - in principle -
enough to determine $R$ by using standard formulas (see for instance
Murphy's book). To determine $\mu$, one can try to solve for it in the
exactness condition, obtained applying Euler's operator to the total
derivative $H \equiv \mu\l(\yn-\Phi\r)$:

\begin{equation}
{\frac {\pa H}{\pa y}}
-{\frac {d}{dx}}
\l(\frac{\pa H}{\pa \y1}\r)
+{\frac {d^{2}}{d{x}^{2}}}
\l( \frac{\pa H}{\pa \yt} \r)
+ ...
+ (-1)^n
{\frac {d^{n}}{d{x}^{n}}} \l( \frac{\pa H}{\pa y^{(n)}} \r) =0
\label{EC}
\end{equation}

% \begin{equation}
% {{ H}_{y}}
% -{\frac {d}{dx}}
% \l({H}_{y'}\r)
% +{\frac {d^{2}}{d{x}^{2}}}
% \l( {H}_{y''} \r)
% + ...
% + (-1)^n
% {\frac {d^{n}}{d{x}^{n}}}
% \l({H}_{y^{(n)}} \r) =0
% \label{EC}
% \end{equation}

\ni \eq{EC} is of the form

\begin{equation}
A(x,y,\y1,...,y^{(2n-3)}) + y^{(2n-2)} B(x,y,\y1,...,\ym) = 0
\label{PDE_ABn}
\end{equation}

\ni where $A$ is of degree $n-1$ in $\yn$ and linear in $\yk$ for $n < k
\leq (2n-3)$, so that \eq{PDE_ABn} can be split into a PDE system for
$\mu$. In the case of second order ODEs - the subject of this work -
\eq{EC} is of the form

\begin{equation}
A(x,y,\y1)+  \yt\ B(x,y,\y1)=0
\label{PDE_AB}
\end{equation}

\ni and the PDE system is obtained by taking $A$ and
$B$ equal to zero\footnote{In a recent work by
\cite{bluman1}, the authors arrive at \eq{PDE_AB} and
\eq{PDE_B} departing from the
adjoint linearized system corresponding to a given ODE; the possible
splitting of \eq{PDE_ABn} into an overdetermined system for $\mu$ is also
mentioned. However, in that work, $\yt$ of \eq{PDE_AB}
above appears replaced by $\Phi(x,y,\y1)$, and the authors discuss
possible alternatives to tackle Eqs.(\ref{PDE_AB}) and (\ref{PDE_B})
instead of Eqs.(\ref{PDE_A}) and (\ref{PDE_B}).}:

\begin{eqnarray}
A  & \equiv &
% \left ({\frac {\pa ^{2}\mu}{\pa \y1\pa x}}
% + \y1{\frac {\pa^{2}\mu}{\pa \y1\pa y}}
% -{\frac {\pa \mu}{\pa y}}\right)\Phi
% +\left ({\frac { \pa^{2}\Phi}{\pa \y1\pa x}}
% -{\frac {\pa \Phi}{\pa y}}
% +\y1{\frac { \pa ^{2}\Phi}{\pa \y1\pa y}}\right )\mu
% +{\y1}^{2}{\frac {\pa ^{2}\mu}{ \pa {y}^{2}}}
% 
\left (y'\mu_{{y'y}}-\mu_{{y}}+\mu_{{y'x}}\right )\Phi+\left (
\Phi_{{y'x}}+y'\Phi_{{y'y}}-\Phi_{{y}}\right )\mu+{y'}^{2}\mu_{{
yy}}
\nonumber
\\*[.1in]
& &
+\left (\mu_{{y}}\Phi_{{y'}}+\mu_{{y'}}\Phi_{{y}}+2\,\mu_{{xy}
}\right )y'+\mu_{{y'}}\Phi_{{x}}+\mu_{{x}}\Phi_{{y'}}+\mu_{{xx}}
= 0
\label{PDE_A}
\\*[.2in]
B
    & \equiv &
% 2\,{\frac {\pa \mu}{\pa y}}
% +\Phi\;{\frac { \pa ^{2}\mu}{\pa {\y1}^{2}}}
% +2 \,{\frac {\pa \mu}{\pa \y1}}\;{\frac {\pa \Phi}{\pa \y1}}
% +\mu{\frac {\pa ^{ 2}\Phi}{\pa {\y1}^{2}}}
% +{\frac {\pa ^{2}\mu}{\pa \y1\pa x}}
% +\y1{\frac {\pa^{2}\mu}{\pa \y1 \pa y}}
y'\mu_{{y'y}}+\Phi\,\mu_{{y'y'}}+\mu\,\Phi_{{y'y'}}+2\,\mu_{{y}}
+2\,\mu_{{y'}}\Phi_{{y'}}+\mu_{{y'x}}
= 0 
\label{PDE_B}
\end{eqnarray}

\ni Regarding the solvability of these equations, unless a more concrete
ansatz for $\mu(x,y,\y1)$ is given, the problem is in principle as
difficult as solving the original ODE. We then studied the solution for
$\mu$ of Eqs.(\ref{PDE_A}) and (\ref{PDE_B}) when $\mu$ depends only on
two variables, that is: for $\mu(x,y)$, $\mu(x,\y1)$ and $\mu(y,\y1)$.
Concretely, we searched for the existence conditions for such integrating
factors, expressed as a set of equations in $\Phi$, plus an algebraic
expression for $\mu$ as a function of $\Phi$, valid when the existence
conditions hold. Formulating the problem in that manner and taking into
account the integrability conditions of the system, Eqs.(\ref{PDE_A}) and
(\ref{PDE_B}) turned out to be solvable for $\mu(x,y)$, but appeared to us
untractable when $\mu$ depends on two variables one of which is $\y1$.

We then considered a different approach, taking into account from the
beginning the form of the ODE family admitting a given integrating factor.
As shown in the following sections, it turns out that, using that piece of
information (\eq{ODE_mu} below), when $\mu$ depends only on two variables
the existence conditions and the integrating factors themselves can be
systematically determined; and in the cases $\mu(x,\y1)$ and $\mu(y,\y1)$,
this can be done without solving any differential equations.

Concerning the ODE families admitting given integrating factors, we note
that, from \eq{exact_oh}

\begin{equation}
\mu(x,y,\y1,...,\ym)=\fr{\pa\,R\mbox{\hspace{2mm}}}{\pa \ym}
\end{equation}

\ni and hence the first integral $R$ is of the form

\begin{equation}
R=G(x,y,...,\ymm)+\int \mu\ d\ym
\label{R}
\end{equation}

\ni for some function $G$. In turn, since $R$ is a first integral, it
satisfies

\begin{equation}
% \frac{\pa R}{\pa x}
% + \y1 \frac{\pa R }{\pa y}
% + \Phi \frac{\pa R}{\pa \y1} = 0
% 
R_x + \y1 R_y + ... + \Phi R_{y^{(n-1)}} = 0
\label{A}
\end{equation}

\ni Inserting \eq{R} into the above and solving for $\yn$ leads to
the general form of an ODE admitting a given integrating factor:

\begin{equation}
\label{ODE_mu}
\yn = \frac {-1}{\mu}
\l[{
    \frac {\pa }{\pa x}\l(\int \mu \, {d{\it \ym}} + G\r)
    + ...
    + \ym \frac {\pa }{\pa  \ymm}\l(\int \mu \, d\ym + G\r)
}
\r]
\end{equation}

% \ni The expression above then becomes an {\it ODE pattern} which one can try
% to match against an input ODE. The generation of such {\it pattern matching
% routines} is difficult, even for restricted subfamilies of integrating
% factor, but once built, they are a powerful and computationally efficient
% way to reduce the order of the corresponding ODEs (see \se{tests}).

\subsection{Second order ODEs and integrating factors of the form
$\mu(x,y)$} \label{mu_xy}

We consider first the determination of integrating factors of the form
$\mu(x,y)$, which turns out to be straightforward\footnote{The result for
\underline{Case A} presented in this subsection is also presented
as lemma 3.8 in \cite{sheftel}.}. The determining
equations (\ref{PDE_A}) and (\ref{PDE_B}) for this case are given by:

\begin{equation}
\begin{array}{c}
{y' }^{2}\mu_{{yy}}+2\,\mu_{{xy}}y' +\mu\,\Phi_{{y' x}}+\mu\,\Phi_{
{y' y}}y' -\mu\,\Phi_{{y}}-\mu_{{y}}\Phi+\mu_{{y}}y' \,\Phi_{{y' }}+
\mu_{{xx}}+\mu_{{x}}\Phi_{{y' }} = 0
\\*[.1in]
\mu\,\Phi_{{y' y' }}+2\,\mu_{{y}}=0
\end{array}
\label{sys_mu_xy}
\end{equation}

\ni Although the use of integrability conditions is enough to tackle this
problem, the solving of Eqs.(\ref{sys_mu_xy}) can be directly simplified
if we take into account the ODE family admitting an integrating factor
$\mu(x,y)$. From \eq{ODE_mu}, that ODE family takes the form

\begin{equation}
\label{mu_xy_ode}
\yt = a(x,y)\, \y1 ^2 + b(x,y)\, \y1 + c(x,y),
\end{equation}

\ni where

\begin{equation}
\label{mu_xy_ode_cond}
a(x,y) = - \frac{\mu_y}{\mu},\ \ \ 
b(x,y) = - \frac{G_y + \mu_x}{\mu},\ \ \ 
c(x,y) = - \frac{G_x}{\mu}
\end{equation}

\ni and $G(x,y)$ is an arbitrary function of its arguments. Hence, as a
shortcut to solving Eqs.(\ref{sys_mu_xy}), one can take \eq{mu_xy_ode} as
an existence condition - $\Phi$ must be a polynomial of degree two in \y1
- and directly solve Eqs.(\ref{mu_xy_ode_cond}) for $\mu$. The
calculations are straightforward; there are two different cases.

\medskip
\ni {\underline {Case A}}: $\ 2a_x-b_y \neq 0$
\medskip

\ni Defining the two auxiliary quantities

\begin{equation}
\varphi \equiv c_y - a\,c - b_x,\ \ \ \ \
\Upsilon \equiv a_{xx} + a_x\,b + \varphi_y
\label{auxiliary}
\end{equation}

\ni an integrating factor of the form $\mu(x,y)$ exists only when

\begin{equation}
\Upsilon_y - a_x = 0,\ \ \ \ \ \
\Upsilon_x + \varphi + b\,\Upsilon - \Upsilon^2 = 0
\end{equation}

\ni and is then given in solved form, in terms of $a$, $b$ and $c$ by

\begin{equation}
\mu(x,y)=
\exp\l(
\int \l(-\Upsilon+ \fr{\pa}{\pa x} \int a\, dy \r) dx
- \int a\, dy
\r)
\end{equation}

\ni So, in this case, when an integrating factor of this type exists there
is only one\footnote{We recall that if $\mu$ is an integrating factor
leading to a first integral $\psi$, then the product $\mu F(\psi)$ - where
$F$ is an arbitrary function - is also an integrating factor, which
however does not lead to a first integral independent of $\psi$.} and it
can be determined without solving any differential equations.

\medskip
\ni {\underline {Case B}}: $\ 2a_x-b_y = 0$
\medskip

\ni Redefining $\varphi \equiv c_y - a\,c$, an integrating factor of the
form $\mu(x,y)$ exists only when

\begin{equation}
a_{xx} - a_x\,b - \varphi_y = 0,
\end{equation}

\ni and then $\mu(x,y)$ is given by

\begin{equation}
\mu(x,y)=\nu(x)\ {\rm e}^{{^{-\displaystyle \int  a\, dy}}}
\label{mu_xy_2}
\end{equation}

\ni where $\nu(x)$ is either one of the independent solutions of the second
order linear ODE\footnote{When the given ODE is linear, \eq{lode} is
just the corresponding adjoint equation.}

\begin{equation}
\nu'' = A(x)\, \nu' + B(x)\, \nu,
\label{lode}
\end{equation}

\ni and
\begin{equation}
A(x) \equiv 2\, {\cal I} - b,\ \ \ \ \ \ 
B(x) \equiv
\varphi
+ \l(
{\cal I} - \fr{\pa}{\pa x}
\r)
\l(
    b- {\cal I}
\r),\ \ \ \ \ \ 
{\cal I} \equiv \fr{\pa}{\pa x} \int a\, dy
\end{equation}

\ni So in this case, to transform \eq{mu_xy_2} into an explicit expression
for $\mu$ we first need to solve a second order linear ODE. When the
attempt to solve \eq{lode} is successful, using each of its two
independent solutions as integrating factors leads to the general solution
of \eq{mu_xy_ode}, instead of just a reduction of order.
 
\subsection{Second order ODEs and integrating factors of the form
$\mu(x,\y1)$} \label{theorem}

When the integrating factor is of the form $\mu(x,\y1)$, the determining
equations (\ref{PDE_A}) and (\ref{PDE_B}) become

\begin{equation}
\begin{array}{c}
\left (\Phi_{{y}}y' +\Phi_{{x}}\right )\mu_{{y' }}+
\left (-\Phi_{{y}}+\Phi_{{y' x}}+\Phi_{{y' y}}y' \right )\mu+\mu_{{x
x}}+\mu_{{x}}\Phi_{{y' }}+\mu_{{y' x}}\Phi=0
\\*[.1in]
\Phi\,\mu_{{y' y' }}+\mu\,\Phi_{{y' y' }}+2\,\mu_{{y' }}\Phi_{{y' }
}+\mu_{{y' x}}=0
\end{array}
\label{sys_mu_x_y1}
\end{equation}

\ni As in the case $\mu(x,y)$, the solution we are interested in is an
expression for $\mu(x,\y1)$ in terms of $\Phi$, as well as existence
conditions for such an integrating factor expressed as equations in
$\Phi$. However, differently than the case $\mu(x,y)$, we didn't find a
way to solve the $\mu(x,\y1)$ problem just using integrability conditions,
neither working by hand nor using the specialized computer algebra
packages {\it diffalg} \cite{boulier} and {\it standard\_form}
\cite{reid}. We then considered approaching the problem as explained in
the previous subsection, departing from the form of \eq{R} for $\mu =
\mu(x,y')$:

% \footnote{The Crack package \cite{wolf} is not designed to work with an
% arbitrary $\Phi$ and hence cannot be used for this purpose.}

\begin{equation}
\label{reducible1}
\yt = \Phi(x,y,\y1) \equiv
- \frac{F_x + G_x + G_y\, \y1}{F_{y'}} 
\end{equation}

\ni where $G(x,y)$ and $F(x,\y1)$ are arbitrary functions of their
arguments and 

\begin{equation}
\label{mu_F}
{\mu}(x,\y1) = F_{y'}
\end{equation}

\ni Now, \eq{reducible1} is not polynomial in either $x$, $y$ or \y1, and
hence its use to simplify and solve the problem is less straightforward
than in the case $\mu(x,y)$. However, in \eq{reducible1}, all the
dependence on $y$ comes from $G(x,y)$ in the numerator, and as it is shown
below, this fact is a key to solving the problem. Considering ODEs for
which $\Phi_y \neq 0$\footnote{ODEs {\it missing y} may also have
integrating factors of the form $\mu(x,\y1)$. Such an ODE however can
always be reduced to first order by a change of variables, so that the
determination of a $\mu(x,\y1)$ for it is equivalent to solving a first
order ODE problem - not the focus of this work.}, the approach we used can
be summarized in the following three lemmas whose proofs are developed
separately for convenience.

\medskip \ni {\bf{Lemma 1.}} {\it For all linear ODEs of the family
\eq{reducible1}, an integrating factor of the form $\mu(x,\y1)$ such that
$\mu_{y'}\neq 0$, when it exists, can be determined directly from the
coefficient of $y$ in the input ODE}.

\medskip \ni {\bf{Lemma 2.}} {\it For all non-linear ODEs of
\eq{reducible1}, the knowledge of $\mu(x,\y1)$ up to a factor depending on
$x$, that is, of $\F(x,\y1)$ satisfying

\begin{equation}
{\F}(x,\y1)= \frac {\mu(x,\y1)}{\mut(x)}
\label{cal_F}
\end{equation}

\ni is enough to determine $\mut(x)$ by means of an integral.}

\medskip \ni {\bf{Lemma 3.}} {\it For all non-linear ODEs members of
\eq{reducible1}, it is always possible to determine a function
${\F}(x,\y1)$ satisfying \eq{cal_F}.}

% \medskip \ni {\bf{Corollary.}} {\it For all second order ODEs such that
% $\Phi_y \neq 0$, the determination of $\mu(x,\y1)$ (if the ODE is linear
% we assume $\mu_{y'} \neq 0$), when it exists, can be performed
% systematically, without solving any differential equations, and without
% using differential Groebner basis techniques.}

\medskip \ni {\bf{Corollary.}} {\it For all second order ODEs such that
$\Phi_y \neq 0$, the determination of $\mu(x,\y1)$ (if the ODE is linear
we assume $\mu_{y'} \neq 0$), when it exists, can be performed
systematically and without solving any differential equations.}

\subsubsection{Proof of {\bf Lemma 1}}

For \eq{reducible1} to be linear and not missing $y$, either $G_x$ or
$G_y$ must be linear in $y$. Both $G_x$ and $G_y$ cannot simultaneously be
linear in $y$ since, in such a case, $G_x/F_{y'}$ or $y'G_y/F_{y'}$ would
be non-linear in $\{y,y'\}$\footnote{We are only interested in the case
$\mu_{y'}=F_{y'y'}\neq 0$.}; therefore, either $G_{yy}=0$ or $G_{xy}=0$.

\medskip
\ni {\underline {Case A}}: $G_y$ is linear in $y$ and $G_{xy} = 0$
\medskip

Hence, $G$ is given by

\begin{equation}
G = C_2 y^2 + C_1 y + g(x)
\end{equation}

\ni where $g(x)$ is arbitrary. From \eq{reducible1}, in order to have
$y'G_y/F_{y'}$ linear in $\{y,y'\}$, $F_{y'}$ must of the form $\nu(x) y'$
for some function $\nu(x)$. Also, $F_x/F_{y'}$ can have a term linear in
$y'$, and a term proportional to $1/\y1$ to cancel with the one coming
from $G_x/F_{y'}=g'/F_{y'}$, so that

\begin{equation}
F_{y'} = \nu \y1,\ \ \ \ \ F_{x} = \frac{\nu' \y1^2}{2} - g'
\end{equation}

% \footnote{If $\nu'=0$ then $\mu(x,\y1) =
% \y1$ and the ODE (\eq{lode_mu_x_y1}) is linear with constant
% coefficients.}

\ni where the coefficient $\nu'/2$ in the second equation above arises
from the integrability conditions between both equations. \eq{reducible1}
is then of the form

\begin{equation}
\yt=
    -{\frac {{\nu}'}{2\;\nu}}\;\y1
    -{\frac {2\;C_2\,}{\nu}}\;y
    -{\frac {C_1}{\nu}}
\label{lode_mu_x_y1}
\end{equation}

\ni and hence, a linear ODE $\yt = a(x) \y1 + b(x) y$ has an integrating
factor $\mu(x,\y1) = \y1/b$ when $b'/b - 2a = 0$. $\triangle$

\medskip
\ni {\underline {Case B}}: $G_x$ is linear in $y$ and $G_{yy} = 0$
\medskip

\ni In this case, in order to have \eq{reducible1} linear, $F_{y'}$ cannot
depend on $y'$, so that the integrating factor is of the form $\mu(x)$ and
hence the case is of no interest: we end up with the standard search for
$\mu(x)$ as the solution to the adjoint of the original linear ODE.
$\triangle$

%-----------------------------------------------------------

% \medskip
% \ni {\it Example:} Kamke's ODE 37.
% \medskip
% 
% \begin{equation}
% \yt =  -2\,y\, \y1 - f(x)\left (\y1+{y}^{2}\right ) + g(x)
% \label{k37}
% \end{equation}
% 
% \ni This example is interesting\footnote{For ODE 6.37, Kamke shows a
% reduction of order to a general Riccati ODE, based on the theory for ODEs
% having solutions with no movable critical points - see
% \cite{painleve}, and \cite{ince} p 331.} because it has no point
% symmetries for arbitrary $f(x)$ and $g(x)$ (see \se{mu_and_X}). For this
% ODE, $\F(x,\y1)$ was determined (see \se{lemma}) as:
% 
% \begin{equation}
% \F(x,\y1) = 1
% \end{equation}
% 
% \ni from which (\eq{G_eq})
% 
% \begin{equation}
% \frac {G_{y\,x}(x,y) + G_{y\,y}(x,y)\, \y1}{\mut(x)} = -2 y\, f(x) -2\y1
% \end{equation}
% 
% \ni and then as in \eq{varphi_1_2} we obtain
% 
% \begin{eqnarray}
% \varphi_1 & = & -2 y\, f(x)
% \nonumber\\*[.07 in]
% \varphi_2 & = & -2
% \end{eqnarray}
% 
% \ni Using this in \eq{mut_2}, we get
% 
% \begin{equation}
% \mut(x) = {e^{^{\ \displaystyle {\int \! {f(x)}\, {dx}} }}}
% \label{mu_k37}
% \end{equation}
% 
% \ni and so, from \eq{cal_F}, since $\F(x,\y1)=1$, $\mu(x,\y1)=\,\mut(x)$.
% 

\subsubsection{Proof of {\bf Lemma 2}}

It follows from Eqs.(\ref{reducible1}) and (\ref{mu_F}) that, given $\F$
satisfying (\ref{cal_F}), 

\begin{equation}
\frac {\pa}{\pa y}
\medl(\Phi(x,y,\y1)\ \F(x,\y1)\medr) =
- \frac {G_{y\,x}(x,y) + G_{y\,y}(x,y)\, \y1}{\mut(x)}
\label{G_eq}
\end{equation}

\ni Hence, by taking coefficients of \y1 in the above,

\begin{eqnarray}
\varphi_1 & \equiv &
\Phi_y(x,y,\y1)\, \F(x,\y1) -\y1\, \frac{\pa}{\pa \y1}
\medl( \Phi_y(x,y,\y1)\, \F(x,\y1) \medr)
= - \frac {G_{y\,x}(x,y)}{\mut(x)}
\nonumber\\*[.07 in]
\varphi_2 & \equiv &
\frac{\pa}{\pa \y1}
\medl( \Phi_y(x,y,\y1)\, \F(x,\y1) \medr)
= - \frac {G_{y\,y}(x,y)}{\mut(x)}
\label{varphi_1_2}
\end{eqnarray}

\ni where the left-hand-sides can be calculated explicitly
since they depend only on $\Phi$ and the given $\F$. Similarly, 

\begin{eqnarray}
\varphi_3 & \equiv &
-\frac {\pa}{\pa \y1} \medl( \Phi(x,y,\y1)\ \F(x,\y1) \medr)
=
\frac {F_{y'x}(x,\y1) + G_{y}(x,y)}{\mut(x)}
\nonumber\\*[.07 in]
\varphi_4 & \equiv & \frac {\pa}{\pa \y1} \F(x,\y1)
= \frac {F_{y'y'}(x,\y1)}{\mut(x)} 
\label{varphi_3_4}
\end{eqnarray}

\ni Now, since in this case the ODE family \eq{reducible1} is nonlinear by
hypothesis, either $\varphi_2$ or $\varphi_4$ is different from zero, so
that at least one of the pairs $\{\varphi_1,\ \varphi_2\}$ or
$\{\varphi_3,\ \varphi_4\}$ can be used to determine $\mut(x)$ as the
solution of a first order linear ODE. For example, if $\varphi_2 \neq 0$,

\begin{equation}
\frac {\pa}{\pa y} \l(\varphi_1(x,y)\ {\mut(x)} \r)
= \frac {\pa}{\pa x} \l(\varphi_2(x,y)\ {\mut(x)} \r)
\end{equation}

\ni from where

\begin{equation}
\mut(x)  = 
{e^{^{\ \displaystyle
    {\int \!
        {\frac {1}{\varphi_2}}
        \l(
            {{\frac {\pa \varphi_1}{\pa y}}}
            -{{\frac {\pa \varphi_2}{\pa x}}}
        \r)
    {dx}}
}}}
\label{mut_2}
\end{equation}

\ni If $\varphi_2 = 0$ then $\varphi_4 \neq 0$ and we obtain

\begin{equation}
\mut(x)  = 
{e^{^{\ \displaystyle
    {\int \!
        {\frac {1}{\varphi_4}}
        \l(
            {{\frac {\pa \varphi_3}{\pa \y1}}}
            -{{\frac {\pa \varphi_4}{\pa x}}}
        \r)
    {dx}}
}}}
\label{mut_3}
\end{equation}

\ni When combined with \eq{cal_F}, \ni Eqs.(\ref{mut_2}) and (\ref{mut_3})
alternatively give both an explicit solution to the problem and an
existence condition, since a solution $\mut(x)$ - and hence an integrating
factor of the form $\mu(x,\y1)$ - exists if the integrand in \eq{mut_2} or
\eq{mut_3} only depends on $x$. $\triangle$

\subsubsection{Proof of {\bf Lemma 3}}
\label{lemma}

We start from \eq{reducible1} by considering the expression

\begin{equation}
\label{Upsilon}
\Upsilon \equiv 
\Phi_y =
- \frac {G_{x\,y}(x,y) + G_{y\,y}(x,y)\ \y1}{F_{y'}(x,\y1)}
\end{equation}

\ni and develop the proof below splitting the problem into different
cases. For each case we show how to find $\F(x,\y1)$ satisfying
\eq{cal_F}. $\F$ will then lead to the required integrating factor when,
in addition to the conditions explained below, the existence conditions
for $\mut(x)$ mentioned in the previous subsection are satisfied.

\smallskip
\ni {\underline{Case A}}: $G_{x\,y} / G_{y\,y}$ depends on
$y$
\smallskip

To determine whether this is the case, we cannot just analyze the ratio
$G_{x\,y}/G_{y\,y}$ itself since it is unknown. However, from
\eq{Upsilon}, in this case the factors of $\Upsilon$ depending on $y$ will
also depend on \y1, and this condition can be formulated as

\begin{equation}
% \Upsilon_y \neq 0 \ \ \mbox{and}\ \ \
\frac{\partial }{\partial \y1} \l(\frac{\Upsilon_y}{\Upsilon}\r) \neq 0
\label{conditions_A}
\end{equation}

\ni When this inequation holds, we determine $F_{y'}(x,\y1)$ up to a
factor depending on $x$, that is, the required $\F(x,\y1)$, as the
reciprocal of the factors of $\Upsilon$ which depend on \y1 but not $y$.
$\triangle$

%%%%%%%%%%%%%%%%%%%%%%%%%

\medskip
\ni {\it Example:} Kamke's ODE 226
\medskip

\ni This ODE is presented in Kamke's book already in exact form, so we start
by rewriting it in explicit form as

\begin{equation}
\yt={\frac {x^2 y\y1+x y^2}{\y1}}  % this was ode 226
% y\yt-{\y1}^{2}+\left (x+x{y}^{2}\right )\y1-y\left (1-{y}^{2}\right ) = 0
% above is ode[123]
\label{k226}
\end{equation}

\ni We determine $\Upsilon$ (\eq{Upsilon}) as 

\begin{equation}
\Upsilon = \frac {x (x \y1+2 y)}{\y1}
\end{equation}

\ni The only factor of $\Upsilon$ containing $y$ is:

\begin{equation}
x \y1+2 y
\end{equation}

\ni and since this also depends on \y1, $\F(x,\y1)$ is given by

\begin{equation}
\F(x,\y1) = \y1
\end{equation}

%%%%%%%%%%%%%%%%%%%%%%%%%

\smallskip
\ni {\underline{Case B}}: either $G_{xy}=0$ or $G_{yy}=0$
\smallskip

When the expression formed by all the factors of $\Upsilon$ containing $y$
does not contain \y1, in \eq{conditions_A} we will have
$\frac{\partial}{\partial y'}(\frac{\Upsilon_y}{\Upsilon}) = 0$, and it is
impossible to determine {\it a priori} whether one of the functions
$\{G_{x\,y},\ G_{y\,y}\}$ is zero, or alternatively their ratio
does not depend on $y$. We then proceed by assuming the former, build an
expression for $\F(x,\y1)$ as in Case A, and check for the existence of
$\mut(x)$ as explained in the previous subsection. If $\mut(x)$ exists,
the problem is solved; otherwise we proceed as follows.

\medskip
\ni {\underline{Case C}}: $G_{x\,y} / G_{y\,y}\neq 0 $ and
does not depend on $y$
\smallskip

In this case, neither $G_{x\,y}$ nor $G_{y\,y}$ is zero and their
ratio is a function of just $x$, so that

\begin{eqnarray}
G_{x\,y} & = &v_1(x)\ w(x,y)
\nonumber\\*[.07 in]
G_{y\,y} & = &v_2(x)\ w(x,y)
\label{rho_v}
\end{eqnarray}

\ni for some unknown functions $v_1(x)$ and $v_2(x)$. \eq{Upsilon} is then
given by

\begin{equation}
\Upsilon = w(x,y)\ \frac{\l(v_1(x) + v_2(x)\ \y1\r)}{F_{y'}(x,\y1)}
\label{case_C}
\end{equation}

%($v_{y'} \neq 0$)

\ni for some function $w(x,y)$, which is made up of the factors of
$\Upsilon$ depending on $y$ and not on \y1. To determine $F_{y'}(x,\y1)$
up to a factor depending on $x$, we need to determine the ratio
$v_1(x)/v_2(x)$. For this purpose, from \eq{rho_v} we build a PDE for
$G_y(x,y)$,

\begin{equation}
\label{case_C_pde}
G_{x\,y} = \frac {v_1(x)}{v_2(x)}\, G_{y,y}
\end{equation}

\ni The general solution of \eq{case_C_pde} is

\begin{equation}
\label{calG}
G_y = \G\l(y + p(x)\r)
\end{equation}

\ni where $\G$ is an arbitrary function of its argument and for
convenience we introduced

\begin{equation}
p'(x)\equiv v_1(x)/v_2(x)
\end{equation}

\ni We now determine $p'(x)$ as follows. Taking into account
\eq{rho_v},

\begin{equation}
\label{case_1_G_2}
v_2(x)\, w(x,y) = \G'(y+p(x))
\end{equation}

\ni By taking the ratio between this expression and its derivative w.r.t 
$y$ we obtain

\begin{equation}
\H(y+p(x))\equiv
\frac {\pa \ln(w)}{\pa y}
= \frac {\G''(y + p(x))}{\G'(y + p(x))}
\label{cal_H} 
\end{equation}

\ni that is, a function of $y+p(x)$ only, which we can determine since we
know $w(x,y)$. If $\H' \neq 0$, $p'(x)$ is given by

\begin{equation}
\label{p_prime}
p'(x)  =  \frac{\H_x}{\H_y}
    =  
\frac {w_{xy} w - w_x w_y } {w_{yy} w - {w_y}^2 }
\end{equation}

\ni In summary, the conditions for this case are

\begin{equation}
\Upsilon_y \neq 0,\ \ \ 
\frac{\partial }{\partial \y1} \l(\frac{\Upsilon_y}{\Upsilon}\r)= 0,\ \ \
\frac {\pa^2 }{\pa y \pa x} \ln(w) \neq 0,\ \ \ 
\frac {\pa^2 }{\pa y \pa y} \ln(w) \neq 0
\label{conditions_C}
\end{equation}

\ni and then, from \eq{case_C}, $\F(x,\y1)$ is given by

\begin{equation}
\F(x,\y1)= \frac{(p'+\y1)\ w}{\Upsilon}
\label{F_p_w_Upsilon}
\end{equation}

\ni where at this point $\Upsilon$, $w(x,y)$ and $p'(x)$ are all
known. $\triangle$

%%%%%%%%%%%%%%%%%%%%%%%%%

\medskip
\ni {\it Example:} Kamke's ODE 136.
\medskip

We begin by writing the ODE in explicit form as

\begin{equation}
\yt = \frac {h(\y1)}{x-y}
\label{k136}
\end{equation}

\ni This example is interesting since the standard search for point
symmetries is made difficult by the presence of an arbitrary function of
$\y1$. $\Upsilon$ (\eq{Upsilon}) is determined as 

\begin{equation}
\Upsilon = -\frac {h({\it \y1})}{(x-y)^2}
\end{equation}

\ni and $w(x,y)$ as

\begin{equation}
w(x,y) = \frac {1}{(x-y)^2}
\end{equation}

\ni Then $\H(y + p(x))$ (\eq{cal_H}) becomes 

\begin{equation}
\H = \frac {2}{x-y}
\end{equation}

\ni and hence, from \eq{p_prime},  $p'(x)$ is 

\begin{equation}
p'(x) = -1
\end{equation}

\ni so from \eq{F_p_w_Upsilon}:

\begin{equation}
\F(x,\y1)= \frac{1-\y1} {h({\it \y1})}
\end{equation}

\smallskip
\ni \underline{Case D}: $\H=0$
\smallskip

We now discuss how to obtain $p'(x)$ when $\H'(y+p(x))=0$. We consider
first the case in which $\H=0$. Then, $\G''=0$ and 
the condition for this case is

\begin{equation}
\Upsilon_{y}=0
\end{equation}

\ni Recalling \eq{calG}, $G$ is given by

\begin{equation}
G(x,y) = C_1\ (y + p(x))^2 + C_2\ (y + p(x)) + g(x)
\end{equation}

\ni for some function $g(x)$ and some constants $C_1$, $C_2$. From
\eq{reducible1}, $\Phi(x,y,\y1)$ takes the form 

\begin{equation}
\label{case_1_subcase}
\Phi(x,y,\y1) = - \frac {F_x(x,\y1) + g'(x) + (2 C_1\ (y + p(x)) + C_2)
(\y1+p'(x))}{F_{y'}(x,\y1)}
\end{equation}

\ni We now determine $p'(x)$ as follows. First, from the knowledge of
$\Upsilon$ and $\Phi$ we build the two explicit expressions:

\begin{equation}
\label{Lambda1}
\Lambda \equiv \frac {1}{\Upsilon} = - \frac {F_{y'}} {2 C_1\ (\y1+p'(x))}
\end{equation}

\ni and

\begin{equation}
\label{Psi1}
\Psi \equiv \frac {\Phi(x,y,\y1)}{\Upsilon} - y = 
\frac {F_x + g'(x)} {2 C_1\ (\y1+p'(x))} + p(x) + \frac{C_2}{2 C_1}
\end{equation}

\ni From \eq{Lambda1} and \eq{Psi1} $\Lambda$ and
$\Psi$ are related by:

\begin{equation}
\label{case_1_subcase_part2}
\frac {\pa}{\pa x} \medl((\y1+p'(x))\ \Lambda\medr) +
\frac {\pa}{\pa \y1} \medl((\y1+p'(x))\ \Psi\medr) 
= p(x) + \frac {C_2}{2 C_1} 
\end{equation}

\ni where the only unknowns are $p(x)$, $C_1$, and $C_2$. By differentiating
the equation above w.r.t \y1 and $x$ we obtain two equations where the only
unknown is $p'(x)$:

% \ni  Differentiating
% with respect to $\y1$ removes the unknown constants $C_1$ and $C_2$:

\begin{eqnarray}
& \Lambda_{y'} p''(x) + (\Lambda_{x y'} + \Psi_{y' y'}) (\y1+p'(x)) +
\Lambda_x + 2 \Psi_{y'} = 0 &
\label{case_1_subcase_part3}
\\*[.11 in]
& \Lambda\ p'''(x) + (\Lambda_{x x} + \Psi_{y' x}) (\y1 + p'(x))
+ (\Lambda_x + \Psi_{y'}) p''(x) + \Psi_x = p'(x) &
\label{case_1_subcase_part4}
\end{eqnarray}

\ni from where we obtain $p'(x)$ by solving a linear algebraic equation built by
eliminating $p''(x)$ between \eq{case_1_subcase_part3} and
\eq{case_1_subcase_part4}\footnote{From \eq{Lambda1}, $\Lambda \neq 0$, so
that \eq{case_1_subcase_part4} always depends on $p'''(x)$, and solving
\eq{case_1_subcase_part3} for $p''(x)$ and substituting twice into
\eq{case_1_subcase_part4} will lead to the desired equation for $p'(x)$.
If \eq{case_1_subcase_part3} depends on $p'(x)$ but not on $p''(x)$, then
\eq{case_1_subcase_part3} itself is already a linear algebraic equation
for $p'(x)$.}. Also, as a shortcut, if $(\Lambda_{x y'} + \Psi_{y' y'})
/ \Lambda_{y'}$ depends on \y1, then we can build a linear algebraic
equation for $p'(x)$ by solving for $p''(x)$ in \eq{case_1_subcase_part3}
and differentiating w.r.t. \y1. $\triangle$

\medskip \ni {\bf{Remark}} \medskip 

If \eq{case_1_subcase_part3} depends neither on $p'(x)$ nor on $p''(x)$
this scheme will not succeed. However, in that case the original ODE is
actually linear and given by \eq{lode_mu_x_y1}. To see this, we set to
zero the coefficients of $p'(x)$ and $p''(x)$ in
\eq{case_1_subcase_part3}, obtaining:

\begin{equation}
\label{case_1_subcase_part5}
\Lambda_{y'} = \Lambda_{x y'} + \Psi_{y' y'} = \Lambda_x + 2
\Psi_{y'}=0
\end{equation}

\ni which implies that $\Lambda$ is a function of $x$ only, and then

\begin{equation}
\label{case_1_subcase_part6}
\Psi_{y' y'} = 0 
\end{equation}

\ni If we now rewrite $F(x,\y1)$ as

\begin{equation}
\label{case_1_subcase_part7}
F(x,\y1) = Z(x,\y1) -  g(x) - \Lambda (\y1 + p')^2\,C_1
\end{equation}

\ni and introduce this expression in \eq{Lambda1}, we obtain $Z_{y'} =
0$; similarly, using this result, \eq{Psi1}, \eq{case_1_subcase_part6} and
\eq{case_1_subcase_part7} we obtain $Z_x=0$. Hence, $Z$ {\it is a
constant}, and taking into account \eq{case_1_subcase_part7} and
\eq{case_1_subcase}, the ODE which led us to this case is just a
non-homogeneous linear ODE of the form

\begin{equation}
\label{case_1_subcase_part9}
(y+p)''+ (\Lambda' (y+p)' - 2 (y + p) - C_2/C_1) / 2 \Lambda=0
\end{equation}

\ni whose homogeneous part does not depend on $p(x)$:

\begin{equation}
\yt+\frac{\Lambda'(x)}{2{\Lambda(x)}}\,\y1-{\frac {y}{\Lambda(x)}}=0
\end{equation}

\ni and as mentioned, it is the same as \eq{lode_mu_x_y1}.

%%%%%%%%%%%%%%%%%%%%%%%%%

\medskip
\ni {\it Example:} Kamke's ODE 66.
\medskip

This ODE is given by

\begin{equation}
\yt = a\left (c+bx+y\right )\left ({\y1}^{2}+1\right )^{3/2}
\end{equation}

\ni Proceeding as in Case A, we determine $\Upsilon$, $w(x,y)$, and
$\H(y+p(x))$ as

\begin{equation}
\Upsilon = a\left (\y1^{2}+1\right )^{3/2};\ \ \ \ \ \ \ 
w(x,y) = 1;\ \ \ \ \ \ \ 
\H = 0
\end{equation}

\ni From the last equation we realize that we are in Case D. We determine
$\Lambda$ and $\Psi$ (Eqs. (\ref{Lambda1}), (\ref{Psi1})) as:

\begin{eqnarray}
\Lambda & = & \frac{1}{\left ({\y1}^{2}+1\right )^{3/2}\, a}
\nonumber\\*[.07 in]
\Psi & = & c + b\, x
\end{eqnarray}

\ni We then build \eq{case_1_subcase_part2} for this ODE:

\begin{equation}
\frac{p''(x)}{\left ({\y1}^{2}+1\right )^{3/2}\, a} +
c + b\, x = p(x) + \frac {C_2}{2 C_1}
\end{equation}

\ni Differentiating w.r.t. \y1 leads to \eq{case_1_subcase_part3}:

\begin{equation}
-3 \frac{p''(x)\, \y1}{\left ({\y1}^{2}+1\right )^{5/2}\, a} = 0
\end{equation}

\ni from which it follows that $p''(x) = 0$. Using this in
\eq{case_1_subcase_part4} we obtain:

% \begin{equation}
% b + \frac{p'''(x)}{\left ({\y1}^{2}+1\right )^{3/2}\, a} - p'(x) = 0
% \end{equation}
% 
% \ni we obtain

\begin{equation}
p'(x) = b
\end{equation}

\ni after which \eq{F_p_w_Upsilon} becomes

\begin{equation}
\F(x,\y1)= \frac{\y1+b} {a \left ({\y1}^{2}+1\right )^{3/2}}
\end{equation}

%%%%%%%%%%%%%%%%%%%%%%%%%

\smallskip
\ni \underline{Case E}: $\H' = 0$ and $\H\neq 0$
\smallskip

In this case $\H(y+p(x)) = \G''/\G'= C_1$, so $\G'$ is an exponential function
of its argument $(y+p(x))$ and hence from \eq{calG}

\begin{equation}
G(x,y) = C_2 e^{(y+p(x))C_1} +  (y+p(x))C_3 + g(x)
\end{equation}

\ni for some constants $C_2$, $C_3$ and some function $g(x)$. In this 
case one of the conditions to be satisfied is

\begin{equation}
\Upsilon_{y}= \mbox{constant} \neq 0
\end{equation}

\ni and $\Phi(x,y,\y1)$ will be of the form

\begin{equation}
\label{case_1_subcase_0}
\Phi(x,y,\y1) = - \frac {F_x(x,\y1) + g'(x) 
+ \l(C_2 C_1 e^{ (y + p(x))C_1} + C_3\r)
\l(\y1+p'(x)\r)}{F_{y'}(x,\y1)}
\end{equation}

\ni Taking advantage of the fact that we explicitly know $C_1$, we
build a first expression for $p'$ by dividing $C_1 e^{y C_1}$ by
$\Upsilon$:

\begin{equation}
\label{Lambda2}
\Lambda \equiv - \frac {F_{y'}} {C_2 e^{p(x)C_1}\ (\y1+p'(x))}
\end{equation}

\ni We obtain a second expression for $p'$ by multiplying $\Phi$ by
$\Lambda$ and subtracting $C_1 e^{C_1 y}$

\begin{equation}
\label{Psi2}
\Psi \equiv  \frac {1}{C_2 e^{p(x)C_1}} \l(\frac {F_x + g'(x)} {\y1+p'(x)} 
+ C_3\r)
\end{equation}

\ni As in Case D, $\Lambda$ and $\Psi$ are related by

\begin{equation}
{\frac {\pa}{\pa x} \medl((\y1+p'(x))\,\Lambda\medr)
+ \l(\y1+p'(x)\r)\, p'(x) \Lambda C_1
+ \frac {\pa}{\pa \y1} \medl((\y1+p'(x))\,\Psi\medr)}
=\frac {C_3}{C_2 e^{p(x)C_1}}
\label{case_1_subcase_0_part2}
\end{equation}

\ni where the only unknowns are $C_2$, $C_3$ and $p(x)$. Differentiating
\eq{case_1_subcase_0_part2} with respect to $\y1$ we have

% \begin{eqnarray}
% \lefteqn{
% \label{case_1_subcase_0_part3}
% \medl(p''(x) + {p'(x)}^2 C_1\medr) \Lambda_{y'}
%         + p'(x) \medl(\y1 \Lambda_{y'} C_1 + \Lambda C_1
%         + \Lambda_{x \y1} + \Psi_{y' y'}\medr)
%         }
% \\
% &\mbox{\hspace{5.5cm}} & 
% + 2 \Psi_{y'} + \Lambda_x + \y1 \Lambda_{x \y1}
% + \y1 \Psi_{y' y'} = 0
% \nonumber
% \end{eqnarray}

\begin{equation}
\begin{array}{rcl}
\lefteqn{
\label{case_1_subcase_0_part3}
\medl(p''(x) + {p'(x)}^2 C_1\medr) \Lambda_{y'}
        + p'(x) \medl(\y1 \Lambda_{y'} C_1 + \Lambda C_1
        + \Lambda_{x y'} + \Psi_{y' y'}\medr)
        }
\\*[.15in]
&\mbox{\hspace{5.5cm}} & 
+ 2 \Psi_{y'} + \Lambda_x + \y1 \Lambda_{x y'}
+ \y1 \Psi_{y' y'} = 0
\nonumber
\end{array}
\end{equation}

\ni The problem now is that, due to the exponential on the RHS of
\eq{case_1_subcase_0_part2}, differently from Case D, we are not able to
obtain a second expression for $p'(x)$ by differentiating w.r.t $x$. The
alternative we have found can be summarized as follows. We first note that
if $\Lambda_{y'}=0$, \eq{case_1_subcase_0_part3} is already a linear
algebraic equation\footnote{We can see this by assuming that
$\Lambda_{y'}=0$ and that \eq{case_1_subcase_0_part3} does not contain
$p'$, and then arriving at a contradiction as follows. We first set the
coefficients of $p'$ in \eq{case_1_subcase_0_part3} to zero, arriving at

\begin{displaymath}
\label{case_1_subcase_0_part3_b}
\mbox{\hspace{3.5cm}}
0 = C_1\ \Lambda + \Psi_{y' y'} = 2 \Psi_{y'} + \Lambda_x 
+ \Psi_{y' y'}
\y1
\mbox{\hspace{3.5cm}}
{(A)}
\end{displaymath}

\ni Eliminating $\Psi_{y' y'}$ gives
$$2 \Psi_{y'} = C_1\ \Lambda\ \y1 - \Lambda_x$$

\ni Differentiating the expression above w.r.t \y1 and since
$\Lambda_{y'}=0$, we have

$$2 \Psi_{y' y'} = C_1\ \Lambda$$

\ni Finally, using Eq.(A), $0 = \Lambda$,
contradicting $F_{y'} \neq 0$.} for $p'$, so that we are only worried
with the case $\Lambda_{y'} \neq 0$. With this in mind, we divide
\eq{case_1_subcase_0_part3} by $\Lambda_{y'}$
% 
%  \eq{case_1_subcase_0_part3} for
% $p''$ 
% 
% \begin{equation}
% \label{case_1_subcase_0_part4}
% p''(x) + C_1\ {p'(x)}^2
%         + p'(x) \frac {C_1\ \Lambda_{y'} \y1 + C_1\ \Lambda
%         + \Lambda_{x y'} + \Psi_{y' y'}}{\Lambda_{y'}}
%         + \frac {2 \Psi_{y'} + \Lambda_x + \Lambda_{x y'} \y1
%         + \Psi_{y' y'} \y1}{\Lambda_{y'}} = 0
% \end{equation}
% 
% \ni where the only unknown is $p'(x)$, and, {\it if} the expression above
and, {\it if} the resulting expression depends on \y1, we directly obtain
a linear algebraic equation in $p'(x)$ by just differentiating w.r.t \y1.
$\triangle$

%%%%%%%%%%%%%%%%%%%%%%%%%

\medskip \ni {\it Example\footnote{There are no
examples of this type in all of Kamke's set of non-linear second order
ODEs.}:}

\medskip

\begin{equation}
\yt=\frac {\y1\,\left (x\y1+1\right )\left (-2+{e^y}\right )}{\y1 x^2+\y1-1}
\end{equation}

\ni We determine $\Upsilon$, $w(x,y)$, and
$\H(y+p(x))$ as

\begin{equation}
\Upsilon = \frac {\y1 (x \y1 + 1) e^y}{\y1 x^2 + \y1 - 1};\ \ \ \ \ \ \
w(x,y) = e^y;\ \ \ \ \ \ \ 
\H = 1
\end{equation}

\ni From the last equation we know that we are in Case E. We then determine
$\Lambda$ and $\Psi$ as in Eqs. (\ref{Lambda2}) and (\ref{Psi2}):

\begin{eqnarray}
\Lambda & = & \frac {\y1 x^2 + \y1 - 1}{\y1 (x \y1 + 1)}
\nonumber\\*[.07 in]
\Psi & = & -2
\end{eqnarray}

\ni Now, we build \eq{case_1_subcase_0_part2}:

\begin{equation}
\frac  {1}{x \y1+1} 
\left(\left (p''+ {p'}^{2}+ {\y1}^2 \frac{(x p'-1)}{x \y1+1}\right) 
\left (x^2 +1-\frac {1}{\y1} \right)+2\,x p'-2 \right) 
=\frac {C_3}{C_2\, e^p}
\end{equation}

\ni and, differentiating w.r.t. \y1, we obtain 
\eq{case_1_subcase_0_part3}:

\begin{equation}
\frac {2\,x \y1+1-(x^3+x) {\y1}^2}{ {\y1}^2 (x \y1+1)^2}\left (p''+ {p'}^2
\right )+\frac {2\,\y1-1-2\,x+x \y1}{\left (x \y1+1\right )^3}(x p'-1)=0
\end{equation}

\ni Proceeding as explained, dividing by $\Lambda_{y'}$ and differentiating
w.r.t. \y1, we have

\begin{equation}
\frac {\pa}{\pa \y1} \left({\y1}^2 \frac {2\,\y1-1-2\,x+x \y1}{\left (x
\y1+1\right ) (2\,x \y1+1-(x^3+x) {\y1}^2 )}\right)(x p'-1)=0
\end{equation}

\ni Solving for $p'(x)$ gives $p'(x)=1/x$, from which \eq{F_p_w_Upsilon}
becomes:

\begin{equation}
\F(x,\y1)= \l(\y1-\frac{1}{x}\r)\frac {\y1 x^2 + \y1 - 1}{\y1 (x \y1 + 1)}
\end{equation}

%%%%%%%%%%%%%%%%%%%%%%%%%

\smallskip
\ni {\underline{Case F}}
\smallskip

The final branch occurs when \eq{case_1_subcase_0_part3} divided by
$\Lambda_{y'}$ does not depend on $\y1$ (so that we will not be able to
differentiate w.r.t \y1). In this case we can build a linear algebraic
equation for $p'(x)$ as follows. Let us introduce the label
$\beta(x,p',p'')$ for \eq{case_1_subcase_0_part3} divided by
$\Lambda_{y'}$, so that \eq{case_1_subcase_0_part3} becomes:

\begin{equation}
\label{Lambda_beta}
\Lambda_{y'}(x,\y1)\ \beta(x,p',p'') = 0
\end{equation}

\ni Since we obtained \eq{case_1_subcase_0_part3} by differentiating
\eq{case_1_subcase_0_part2} with respect to \y1,
\eq{case_1_subcase_0_part2} can be written in terms of $\beta$ by
integrating \eq{Lambda_beta} with respect to \y1:

\begin{equation}
\Lambda(x,\y1) \beta(x,p',p'') + \gamma(x,p',p'') = \frac {C_3} {C_2
e^{p(x) C_1}}
\label{Lambda_beta_gamma}
\end{equation}

\ni where $\gamma(x,p',p'')$ is the constant of integration, and can be
determined explicitly in terms of $x$, $p'$ and $p''$ by comparing
\eq{Lambda_beta_gamma} with \eq{case_1_subcase_0_part2}. Taking into account
that ${\beta}(x,p',p'')=0$,  \eq{Lambda_beta_gamma} reduces to:

\begin{equation}
\label{gamma}
\gamma(x,p',p'') = \frac{C_3}{C_2 e^{p(x)C_1}}
\end{equation}

\ni We can remove the unknowns $C_2$ and $C_3$ after multiplying \eq{gamma}
by $e^{p(x)C_1}$, differentiating with respect to $x$, and then dividing
once again by $e^{p(x)C_1}$. We now have our second equation for $p'$, which
we can build explicitly in terms of $p'$, since we know $\gamma(x,p',p'')$
and $C_1$:

\begin{equation}
\label{gamma_p}
\frac {d \gamma}{dx}+C_1\ p' \gamma=0
\end{equation}

\ni Eliminating the derivatives of $p'$ between \eq{Lambda_beta} and 
\eq{gamma_p} leads to a linear algebraic equation in $p'$. Once we have 
$p'$, the determination of $\F(x,\y1)$ follows directly from 
\eq{F_p_w_Upsilon}. $\triangle$

\subsection{Integrating factors of the form $\mu(y,\y1)$}

From \eq{ODE_mu}, the ODE family admitting an integrating factor of the
form $\mu(y,\y1)$ is given by

\begin{equation}
\label{reducible2}
\yt=-{\frac {\y1}{\mu}}
\l(G_{{y}}+\frac{\partial}{\partial y}\int \!\mu\;{d\y1}\r)-{\frac
{G_{{x}}}{\mu}}
\end{equation}

% the following first integral is
% associated with the integrating factor family $\mu(y,\y1)$:
% 
% \begin{equation}
% \label{reduced12}
% R(x,y,\y1)= F(y,\y1)+ G(x,y) 
% \end{equation}
% 
% \ni where $\mu = F_{y'}$, and to this integrating factor corresponds the
% ODE pattern:
% 
% \begin{equation}
% \label{reducible2}
% \yt = - \frac{ G_x(x,y) + (F_y(y,\y1) + G_y(x,y)) \y1} {F_{y'}(y,\y1)}
% \end{equation}

\ni where $\mu(y,y')$ and $G(x,y)$ are arbitrary functions of their
arguments. For this ODE family, it would be possible to develop an
analysis and split the problem into cases as done in the previous section
for the case $\mu(x,\y1)$. However, it is straightforward to notice that
under the transformation $y(x) \rightarrow x,\ x \rightarrow y(x)$,
\eq{reducible2} transforms into an ODE of the form \eq{reducible1} with
integrating factor $\mu(x,\y1^{-1})/\y1^2$. It follows that an integrating
factor for any member of the ODE family above can be found by merely
changing variables in the given ODE and calculating the corresponding
integrating factor of the form $\mu(x,y')$.

\medskip
\ni {\it Example:}
\medskip

\begin{equation}
\yt -{\frac {{\y1}^{2}}{y}} + \sin(x)\,\y1\,y +\cos(x)\,{y}^{2} = 0
\label{gon}
\end{equation}

\ni Changing variables $y(x) \rightarrow x,\ x \rightarrow y(x)$ we
obtain

\begin{equation}
\yt+{\frac {\y1}{x}}-\sin(y)\,{\y1}^{2}x-\cos(y)\,{x}^{2}{\y1}^{3}=0
\label{gon_y_x}
\end{equation}

\ni Using the algorithm outlined in the previous section, an integrating
factor of the form $\mu(x,\y1)$ for \eq{gon_y_x} is given by

\begin{equation}
{\frac {1}{{\y1}^{2}x}}
\end{equation}

\ni from where an integrating factor of the form $\mu(y,\y1)$ for \eq{gon}
is $1/y$, leading to the first integral

\begin{equation}
\sin(x)\;y+{\frac {\y1}{y}}+C_1=0,
\end{equation}

\ni which is a first order ODE of Bernoulli type. The solution to \eq{gon}
then follows directly. This example is interesting since from
\cite{gonzales} \eq{gon} has no point symmetries.

% \subsection{The Connection to PDEs}
% 
% Let $R(x,y,\y1)$ be a first integral of a given second order ODE. We
% rewrite \eq{A} by renaming $\y1\equiv z$
% 
% \begin{equation}
% \label{solvable_PDE}
% R_x + z R_y + \Phi(x,y,z) R_{z} = 0
% % \frac {\pa  R} {\pa  x} + z \frac {\pa  R} {\pa  y}
% % + \Phi(x,y,z) \frac {\pa  R} {\pa  z} = 0
% \end{equation}
% 
% \ni From the results of the previous subsections, if a given PDE is of the
% form \eq{solvable_PDE} with $\Phi_y \neq 0$ and $\Phi$ is nonlinear in
% $\{y,z\}$, then, a particular solution of the form $R(x,y,z) = F(x,z) +
% G(x,y)$, or $R(x,y,z) = F(y,z) + G(x,y)$, when it exist, it can be
% determined in a systematic manner. It is worth mentioning that the
% determination of R in this manner does not require solving the
% characteristic strip of \eq{solvable_PDE}, thus being a concrete
% alternative.

\section{Integrating factors and symmetries}
\label{mu_and_X}

Besides the formulas for integrating factors of the form $\mu(x,y)$, the
main result presented in this paper is a systematic algorithm for the
determination of integrating factors of the form $\mu(x,\y1)$ and
$\mu(y,\y1)$ {\it without solving any auxiliary differential equations or
performing differential Groebner basis calculations}, and these last two
facts constitute the relevant point. Nonetheless, it is interesting to
briefly compare the standard integrating factor ($\mu$) and symmetry
approaches, so as to have an insight of how complementary these methods
can be in practice.

To start with, both methods tackle an {\it n$^{th}$} order ODE by looking
for solutions to a linear {\it n$^{th}$ order determining PDE} in $n+1$
variables. Any given ODE has infinitely many
integrating factors and symmetries. When many solutions to these {\it
determining} PDEs are found, both approaches can, in principle, give a
multiple reduction of order.

In the case of integrating factors there is one unknown function, while for
symmetries there is a pair of infinitesimals to be found. But symmetries are
defined up to an arbitrary function, so that we can always take one of these
infinitesimals equal to zero\footnote{Symmetries
$[\xi(x,y,..y^{(n-1)}),\eta(x,y,..y^{(n-1)})]$ of an \nth order ODE can
always be rewritten as $[G, (G - \xi) y' + \eta]$, where
$G(x,y,..y^{(n-1)})$ is an arbitrary function (for first order ODEs, $y'$
must replaced by the right-hand-side of the ODE). Choosing $G$ = 0 the
symmetry acquires the form $[0,\bar{\eta}]$}; hence we are facing approaches
of equivalent levels of difficulty and actually of equivalent solving power.

Also valid for both approaches is the fact that, unless some {\it
restrictions} are introduced on the functional dependence of $\mu$ or the
infinitesimals, there is no hope that the corresponding determining PDEs
will be easier to solve than the original ODE. In the case of symmetries, it
is usual to restrict the problem to ODEs having {\it point symmetries}, that
is, to consider infinitesimals depending only on $x$ and $y$. The
restriction to the integrating factors here discussed is similar: we
considered $\mu$'s depending on only two variables.

At this point it can be seen that the two approaches are complementary: the
determining PDEs for $\mu$ and for the symmetries are different\footnote{We
are considering here ODEs of order greater than one.}, so that even using
identical restrictions on the functional dependence of $\mu$ and the
infinitesimals, problems which may be untractable using one approach may be
easy or even trivial using the other one.

As an  example of this, consider Kamke's ODE 6.37

% , appearing in this
% paper as \eq{k37}:

\begin{equation}
\yt + 2\,y\, \y1 + f(x)\left (\y1+{y}^{2}\right ) - g(x) = 0
\label{k37}
\end{equation}

\ni For {\it arbitrary} $f(x)$ and $g(x)$, this ODE has an integrating
factor depending only on $x$, easily determined using the algorithms
presented. Now, for {\it non-constant} $f(x)$ and $g(x)$, this ODE has no
point symmetries, that is, no infinitesimals of the form $[\xi(x,y),\
\eta(x,y)]$, except for the particular case in which $g(x)$ can be
expressed in terms of $f(x)$ as in\footnote{To determine $g(x)$ in terms
of $f(x)$ we used the {\it standard form} Maple package by Reid and
Wittkopf complemented with some basic calculations.}

\begin{equation}
g(x)=
\fr{\it f''}{4}\,
+ \fr{3\,f\,{\it f'}}{8}
+ \fr{f^3}{16}
-{\frac {C_2\,{\exp{\l(-3/2\,\displaystyle\int \!f(x){dx}\r)}}}
{4\,\left (2\,C_1+ \displaystyle
\int \!{\exp\l({-1/2\,\int \!f(x){dx}}\r)}\,{dx}\right )^{3}}}
\end{equation}

\ni Furthermore, this ODE does not have non-trivial symmetries of the form
$[\xi(x,\y1),\ \eta(x,\y1)]$ either, and for symmetries of the form
$[\xi(y,\y1),\ \eta(y,\y1)]$ the determining PDE does not split into a
system.

Another ODE example of this type is found in a paper by
\cite{gonzales} (1988):

\begin{equation}
\yt - {\frac {{\y1}^{2}}{y}} - g(x)\,p\,{y}^{p}\y1 - {\it g'}\,{y}^{p+1} = 0
\label{gon2}
\end{equation}

\ni In that work it is shown that for constant $p$, the ODE above only has
point symmetries for very restricted forms of $g(x)$. For instance,
\eq{gon} is a particular case of the ODE above and has no point
symmetries. On the other hand, {\it for arbitrary} $g(x)$, \eq{gon2} has
an obvious integrating factor depending on only one variable: $1/y$,
leading to a first integral of Bernoulli type:

\begin{equation}
{\frac {\y1}{y}}-g(x){y}^{p}+C_1=0
\end{equation}

\ni so that the whole family \eq{gon2} is integrable by quadratures.

We note that \eq{k37} and \eq{gon2} are respectively particular cases of
the general reducible ODEs having integrating factors of the form
$\mu(x)$:

\begin{equation}
\yt=-{\frac {\left (\mu_{{x}}+G_{{y}}\right )}{\mu(x)}}\y1-{\frac {G_{
{x}}}{\mu(x)}}
\end{equation}

\ni where $\mu(x)$ and $G(x,y)$ are arbitrary; and $\mu(y)$:

\begin{equation}
\yt=-{\frac {( \mu_{{y}}\,\y1 + {G_{{y}}}) }{\mu(y)}} \,{\y1}
-{\frac {G_{{x}}}{\mu(y)}}
\end{equation}

\ni In turn, these are very simple cases if compared with the general ODE
families \eq{reducible1} and \eq{reducible2}, respectively having
integrating factors of the forms $\mu(x,\y1)$ and $\mu(y,\y1)$, and which
can be systematically reduced in order using the algorithms here
presented.

It is then natural to conclude that the integrating factor and the
symmetry approaches are useful for solving different types of ODEs, and
can be viewed as equivalently powerful and general, and in practice
complementary. Moreover, if for a given ODE, an integrating factor and a
symmetry are known, in principle one can combine this information to build
two first integrals and reduce the order by two at once (see for instance
\cite{stephani}). 

\section{Tests}
\label{tests}

After plugging the reducible-ODE scheme here presented into the ODEtools
package \cite{odetools2}, we tested the scheme and routines using Kamke's non-linear 246
second order ODE examples\footnote{Kamke's ODEs 6.247 to 6.249 cannot be
made explicit and are then excluded from the tests.}. The purpose was to
confirm the correctness of the returned results and to determine which of
these ODEs have integrating factors of the form $\mu(x,y)$, $\mu(x,\y1)$
or $\mu(y,\y1)$. The test consisted of determining $\mu$ and testing the
exactness condition \eq{EC}.

% ; all the integrating factors found satisfied
% this exactness condition.

In addition, we ran a comparison of performances in solving a subset of
Kamke's examples having integrating factors of the forms $\mu(x,\y1)$ or
$\mu(y,\y1)$, using different computer algebra ODE-solvers (Maple,
Mathematica, MuPAD and the Reduce package Convode). The idea was to
situate the new scheme in the framework of a sample of relevant packages
presently available.

% As a secondary goal, we were also interested in
% comparing the solving performance of the new scheme with the one of the
% symmetry scheme implemented in ODEtools.

% Finally we considered the table of integrating factors for second order
% non-linear ODEs found in Murphy's book and the answers for them returned by
% all these ODE-solvers.

% \subsection{The {\it reducible-ODE} solving scheme and Kamke's ODEs}

To run the comparison of performances, the first step was to classify
Kamke's ODEs into: {\it missing x}, {\it missing y}, {\it exact} and {\it
reducible}, where the latter refers to ODEs having integrating factors of
the forms $\mu(x,\y1)$ or $\mu(y,\y1)$. ODEs missing variables were not
included in the test since they can be seen as first order ODEs in
disguised form, and as such they are not the main target of the algorithm
being presented. The classification we obtained for these 246 ODEs is as
follows

{\begin{center} {\footnotesize
% \begin{tabular}{|p{1.24 in}|p{4.2 in}|}
\begin{tabular}{|p{1.24 in}|p{4.8 in}|}
\hline
Classification  &  ODE numbers as in Kamke's book \\
\hline 
99 ODEs are missing $x$ or missing $y$ &
1, 2, 4, 7, 10, 12, 14, 17, 21, 22, 23, 24, 25, 26, 28, 30, 31, 32, 40,
42, 43, 45, 46, 47, 48, 49, 50, 54, 56, 60, 61, 62, 63, 64, 65, 67, 71,
72, 81, 89, 104, 107, 109, 110, 111, 113, 117, 118, 119, 120, 124, 125,
126, 127, 128, 130, 132, 137, 138, 140, 141, 143, 146, 150, 151, 153, 154,
155, 157, 158, 159, 160, 162, 163, 164, 165, 168, 188, 191, 192, 197, 200,
201, 202, 209, 210, 213, 214, 218, 220, 222, 223, 224, 232, 234, 236, 237,
243, 246 \\
\hline
13 are in exact form &
36, 42, 78, 107, 108, 109, 133, 169, 170, 178, 226, 231, 235
\\
\hline
% 68 ODEs are {\it reducible} with integrating factor 
% $\mu(x,\y1)$ or $\mu(y,\y1)$ &
% 1, 2, 4, 7, 10, 12, 14, 17, 36, 37, 40, 42, 50, 51, 56, 64, 65, 66, 78,
% 81, 89, 97, 104, 107, 108, 109, 110, 111, 123, 125, 126, 133, 134, 135,
% 136, 137, 138, 150, 154, 155, 157, 164, 166, 168, 169, 173, 174, 175, 176,
% 178, 179, 188, 191, 192, 193, 196, 203, 204, 206, 209, 210, 214, 215, 218,
% 220, 222, 226, 235, 236 \\
% 
% 56 ODEs are {\it reducible} with integrating factor 
% $\mu(x,y)$ &
% 36, 37, 40, 42, 51, 78, 79, 84, 93, 97, 98, 107, 108, 109, 113, 117, 122,
% 123, 124, 125, 128, 133, 150, 151, 157, 162, 164, 166, 168, 169, 170, 173,
% 174, 175, 176, 178, 179, 180, 181, 182, 184, 185, 186, 191, 192, 196, 198,
% 203, 204, 206, 210, 214, 215, 216, 218, 222 \\
% \hline
40 ODEs are {\it reducible} with integrating factor 
$\mu(x,\y1)$ or $\mu(y,\y1)$ {\it and}
missing $x$ or $y$&
1, 2, 4, 7, 10, 12, 14, 17, 40, 42, 50, 56, 64, 65, 81, 89, 104, 107, 109,
110, 111, 125, 126, 137, 138, 150, 154, 155, 157, 164, 168, 188, 191, 192,
209, 210, 214, 218, 220, 222, 236 \\
\hline
28 ODEs are {\it reducible} and not missing $x$ or $y$ &
36, 37, 51, 66, 78, 97, 108, 123, 133, 134, 135, 136, 166, 169, 173,
174, 175, 176, 178, 179, 193, 196, 203, 204, 206, 215, 226, 235
\\
% \hline
% 28 ODEs have $\mu(x,\y1)$ or $\mu(y,\y1)$ and not missing $x$ or $y$ &
% 36, 37, 51, 66, 78, 97, 108, 123, 133, 134, 135, 136, 166, 169, 173,
% 174, 175, 176, 178, 179, 193, 196, 203, 204, 206, 215, 226, 235
% \\
\hline
\multicolumn{2}{c}{Table 1. Missing variables, exact and {\it
reducible} Kamke's 246 second order non-linear ODEs.}
\end{tabular}}
\end{center}}

\ni For our purposes, the interesting subset is the one comprised of the
28 ODEs not already missing variables. The results we obtained using the
aforementioned computer algebra ODE-solvers\footnote{Maple R4 is not
present in the table since it is not solving any of these 28 ODEs. This
situation is being resolved in the upcoming Maple R5, where the ODEtools
routines are included in the Maple library, and the previous ODE-solver
was replaced by {\bf odsolve}. However, the scheme here presented was
not ready when the development library was closed; the {\it reducible}
scheme implemented in Maple R5 is able to determine, when they exist,
integrating factors only of the form $\mu(\y1)$.} are summarized as
follows\footnote{Some of these 28 ODEs are given in Kamke in exact form
and hence they can be easily reduced after performing a check for
exactness; before running the tests all these ODEs were rewritten in
explicit form by isolating $\yt$.}:

{\begin{center}
{\footnotesize
\begin{tabular}
{|p{1.2cm}|p{2.2cm}|p{2.6cm}|p{2.6cm}|p{4.7cm}|}
\hline
% & \multicolumn{5}{c} {Kamke's ODE numbers}
\multicolumn{5}{|c|} {Kamke's ODE numbers}
\\
\hline
    & Convode &  Mathematica 3.0 & MuPAD 1.3 &  ODEtools
\\
\hline 
Solved: &
51, 166, 173, 174, 175, 176, 179.
&  
78, 97, 108, 166, 169, 173, 174, 175, 176, 178, 179, 206.
&
78, 97, 108, 133, 166, 169, 173, 174, 175, 176, 179, 
&
51, 78, 97, 108, 133, 134, 135, 136, 166, 169, 173, 174,
175, 176, 178, 179, 193, 196, 203, 204, 206, 215.
\\
\hline 
{\it Totals:} & \centering 7  & \centering 12
& \centering 11 &
{\centering 22

\vspace{-1cm}
}
\\
\hline 
Reduced: &  &  &  & 
36, 37, 66, 123, 226, 235.
\\
\hline 
{\it Totals:} & \centering 0 & \centering 0 & \centering 0 &
{\centering 6

\vspace{-1cm}
}
\\
\hline 
\multicolumn{5}{c}
{Table 2. \it Performances in solving 28 Kamke's ODEs having an integrating
factor $\mu(x,\y1)$ or
$\mu(y,\y1)$}\\
\end{tabular}}
\end{center}}

As shown above, while the scheme here presented is finding first integrals
in all the 28 ODE examples, opening the way to solve 22 of them to the end,
the next scores are only 12 and 11 ODEs, respectively solved by Mathematica
3.0 and MuPAD 1.3.

Concerning the six reductions of order returned by {\bf odsolve}, it must 
be said that neither MuPAD nor Mathematica provide a way to convey them, 
so that perhaps their ODE-solvers are obtaining first integrals for these 
cases but the routines are giving up when they cannot solve the problem to 
the end.

\section{Conclusions}
\label{conclusions}

In connection with second order ODEs, this paper presented a systematic
method for determining the existence of integrating factors and their
explicit form, when they have the forms $\mu(x,y)$, $\mu(x,\y1)$ and
$\mu(y,\y1)$. The scheme is new, as far as we know, and its implementation
in the framework of the computer algebra package ODEtools has proven to be
a valuable tool. Actually, the implementation of the scheme solves ODEs
not solved by using standard or symmetry methods (see \se{mu_and_X}) or
some other relevant and popular computer algebra ODE-solvers (see \se{tests}).

Furthermore, the algorithms presented involve only very simple operations
and do not require solving auxiliary differential equations, except in one
branch of the $\mu(x,y)$ problem. So, even for examples where other
methods also work, for instance by solving the related PDE system
Eqs.(\ref{PDE_A}) and (\ref{PDE_B}) using ansatzes and differential
Groebner basis techniques, the method here presented can return answers
faster and avoiding potential explosions of memory\footnote{Explosions of
memory may happen when calculating all the integrability conditions
involved at each step in the differential Groebner basis approach.}.

% It is also worth mentioning that restricting the dependence of $\mu$ in
% \eq{EC} to $\mu(x,\y1)$, {\it does not lead} to a straightforwardly
% solvable problem. Moreover, when the determination of $\mu$ by directly
% taclking \eq{EC} is frustrated, it is not clear how to determine in
% general whether such a solution $\mu(x,\y1)$ exists.

On the other hand, we have restricted the problem to the universe of
second order ODEs having integrating factors depending only on two
variables while packages as CONLAW (in REDUCE) can try and in some cases
solve the PDE system Eqs.(\ref{PDE_A}) and (\ref{PDE_B}) by using more
varied ansatzes for $\mu$.

% Also, a routine, {\bf redode}, was designed to tackle the inverse problem,
% that is to find the most general \nth order ODE having a given integrating
% factor, optionally reducing to a given $k^{th}$ order ODE ($k<n$). This
% command can be useful in many situations, and for investigating new
% solving methods.

A natural extension of this work would be to develop a scheme for building
integrating factors of restricted but more general forms, now for higher
order ODEs. We are presently working on these possible
extensions\footnote{See http://lie.uwaterloo.ca/odetools.html}, and expect
to succeed in obtaining reportable results in the near future.

\section*{Acknowledgments}

\noindent This work was supported by the State University of Rio de Janeiro
(UERJ), Brazil, and by the Symbolic Computation Group, Faculty of
Mathematics, University of Waterloo, Ontario, Canada.
The authors would like to thank K. von B\"ulow\footnote{Symbolic
Computation Group of the Theoretical Physics Department at UERJ - Brazil.}
for a careful reading of this paper.

\newpage
\appendix
\section*{Appendix A}
% \hspace\parindent

We display here both the integrating factors obtained for the 28 Kamke's
ODEs used in the tests (see \se{tests}) and the ``case" corresponding to
each ODE when using just the algorithms for $\mu(x,\y1)$ or
$\mu(y,y')$\footnote{We note that for non-linear ODEs these two algorithms
work as well when $\mu_{y'} = 0$, but in practice these very simple
examples are covered by the algorithm for $\mu(x,y)$ presented in
\se{mu_xy}.}. As explained in \se{lemma}, the algorithm presented is
subdivided into different cases: A, B, C, D, E and F, and case B is
always either A or C. 

{\begin{center} {\footnotesize
\begin{tabular}{|p{1 in}|p{3 in}|p{.35 in}|}
\hline
Integrating factor  &  Kamke's book ODE-number  & Case
\\
\hline
$1 $ &  36  &  D
\\
\hline
${e^{\int \!f(x){dx}}} $ &  37  &  A
\\
\hline
${{y'}}^{-1} $ &  51, 166, 169, 173, 175, 176, 179, 196, 203, 204, 206, 
                                                               215  &  C
\\
\hline
${\frac {b+{y'}}{\left (1+{{y'}}^{2}\right )^{3/2}}} $ &  66  &  D
\\
\hline
$x $ &  78  &  D
\\
\hline
${x}^{-1} $ &  97  &  A
\\
\hline
$y $ &  108  &  D
\\
\hline
${y}^{-1} $ &  123  &  A
\\
\hline
${\frac {1+{y'}}{\left ({y'}-1\right ){y'}}} $ &  133  &  C
\\
\hline
${\frac {{y'}-1}{\left (1+{y'}\right ){y'}}} $ &  134  &  C
\\
\hline
${\frac {{y'}-1}{\left (1+{y'}\right )\left (1+{{y'}}^{2}\right )}} $ 
                                                             &  135  &  C
\\
\hline
${\frac {{y'}-1}{h({y'})}} $ &  136  &  C
\\
\hline
${\frac {x}{2\,x{y'}-1}} $ &  174  &  C
\\
\hline
$\left (1+{y'}\right )^{-1} $ &  178  &  C
\\
\hline
${\frac {1}{{y'}\,\left (1+2\,y{y'}\right )}} $ &  193  &  C
\\
\hline
${y'} $ &  226  &  A
\\
\hline
$h({y'}) $ &  235  &  C
\\
\hline
\multicolumn{2}{c}
{Integrating factors for Kamke's ODEs which are {\it reducible} and 
not missing $x$ or $y$.}
\end{tabular}}
\end{center}}
\label{lastpage}
\end{document}